\documentclass[structabstract]{aa}
\usepackage{txfonts}

\usepackage{graphicx}
\usepackage{natbib}
\usepackage{longtable}
\usepackage{subeqnarray}
\usepackage{bm}
\usepackage{amsfonts,amssymb}
\usepackage{caption}
\usepackage{algorithm}
\usepackage{enumerate}
\usepackage{nonfloat}
\usepackage{cases}
\usepackage{subfig}
\usepackage{soul}
\usepackage{threeparttable}

\def\st#1{}
\usepackage{color}
\usepackage{amstext}

\begin{document}

\title {Efficient implementation of the adaptive scale pixel decomposition algorithm}

\author{L. Zhang\inst{1,2,3}
 \and S. Bhatnagar\inst{2}
  \and U. Rau\inst{2}
   \and M. Zhang\inst{1,4}}


\institute{Xinjiang Astronomical Observatory, Chinese Academy of Sciences, Urumqi 830011, China \\
 \email{zhangli@xao.ac.cn}
  \and National Radio Astronomy Observatory, Socorro 87801, NM, USA\\
  \email{sbhatnag,rurvashi@nrao.edu}
   \and University of Chinese Academy of Sciences, Beijing 100080, China
    \and Key Laboratory of Radio Astronomy, Chinese Academy of Sciences, Urumqi 830011, China}

\date{Received / Accepted  }

\abstract {Most popular algorithms in use to remove the effects of a
  telescope's point spread function (PSF) in radio astronomy are variants
  of the CLEAN  algorithm. Most of these algorithms model the sky brightness
  using the delta-function basis, which results in undesired artefacts when used on image
  extended emission. The adaptive scale pixel decomposition (Asp-Clean) algorithm models the sky
  brightness on a scale-sensitive basis and thus gives a significantly
  better imaging performance when imaging fields that contain both
  resolved and unresolved emission.}
  {However, the runtime cost of Asp-Clean
  is higher than that of scale-insensitive algorithms. In this
  paper, we identify the most expensive step in the original Asp-Clean
  algorithm and present an efficient implementation of it, which significantly reduces the computational
  cost while keeping the imaging performance comparable to the
  original algorithm.  The PSF sidelobe levels of modern wide-band
  telescopes are significantly reduced, allowing us to make
  approximations to reduce the computing cost, which in turn
  allows for the deconvolution of larger images on reasonable timescales.}
  {As in
  the original algorithm, scales in the image are estimated through
  function fitting.  Here we introduce an analytical method to model
  extended emission, and a modified method for estimating the
  initial values used for the fitting procedure, which ultimately leads to
  a lower computational cost.}
  {The new implementation was tested with
  simulated EVLA data and the imaging performance compared well with the
  original Asp-Clean algorithm. Tests show that the current algorithm
  can recover features at different scales with lower computational cost.}
  {}

\keywords{methods: data analysis -- techniques: image processing}
 \maketitle

\section{Introduction}

Radio interferometers measure the spatial coherence of the electric
field \citep{tho01} in the Fourier domain.  Owing to incomplete
sampling in the Fourier plane, the telescope's PSF has wide-spread
sidelobes. The true sky brightness is convolved with this PSF,
which limits the imaging to no better than a few 100:1 in the dynamic range.

The CLEAN algorithm \citep{hog74} and its variants \citep{cla80, sch83} have so far been
the most popular algorithms in use to remove the effects of the PSF.
Most of these algorithms model the sky brightness as a
collection of delta functions, with an implicit assumption that the
sky is composed of several well-separated point sources. As such, this approach works well for fields
dominated by unresolved sources.  With the increase in sensitivity of
modern radio telescopes, low-level extended emission is often detected
in many fields, particularly at low frequencies. Accurate deconvolution of extended emission is an important problem, especially with the sensitivity of new telescopes. Recently, scale-sensitive methods have been developed to mitigate artefacts induced
by the original CLEAN algorithm when used to deconvolve extended
emission \citep{bha04, cor08, rau11}. These scale-sensitive algorithms have been shown to have better performance than scale-insensitive algorithms in recovering extended emission.

The Multi-Scale CLEAN (MS-Clean) algorithm
\citep{cor08} uses a matched-filtering technique to find the
components. It decomposes a sky image into a set of fixed scales,
such as tapered and truncated paraboloids.
The Asp-Clean algorithm was proposed by \cite{bha04}; here the scales are determined adaptively through a
fitting procedure and heuristics are developed to limit the set of scales (active set) that need to be fitted during any given iteration. Compared to
MS-Clean, the Asp-Clean algorithm gives better imaging performance,
but  at the cost of a significant increase in computing time.

In this paper, we propose an efficient implementation that significantly reduces the computational cost while keeping the imaging
performance comparable to the original Asp-Clean algorithm (referred
to as Asp-Clean2004 in the rest of the text).  This is possible because
the PSF sidelobes of modern wide-band telescopes with excellent
coverage of the uv-plane for continuum imaging are significantly reduced.  This allows us to
approximate the PSF as a single Gaussian. Then the convolution in the objective function of the iterative fitting (using Levenberg -- Marquardt (LM) minimization \citep{mar63}) can be removed. These convolutions are performed through expensive (fast Fourier transformation) FFT-based approach in the Asp-Clean2004 algorithm.  With typically large numbers of
minimization iterations required, FFT-based convolution was responsible
for the high computing cost of the Asp-Clean2004 implementation.  Our current
approach, referred to as Asp-Clean2016 in the text below, can be
thought of as an extreme case of the approximations used in the
Clark CLEAN (Clark-Clean) algorithm \citep{cla80}, where the full PSF is approximated by a PSF-patch
that only includes the highest sidelobe.  As the highest sidelobe becomes increasingly smaller with modern wide-band telescopes, Asp-Clean2016 ignores
the sidelobes of the PSF entirely.  We show that this still leads to
convergence and the total runtime is significantly reduced, even if it is
at the expense of a slightly larger number of iterations.

We recall the basics of the Asp-Clean2004 algorithm in
Sect. $2$, while the Asp-Clean2016 algorithm is described in Sect. $3$. Numerical experiment and comparisons are presented in Sect. $4$, and we summarize our results in Sect. $5$.

\section{Asp-Clean2004 algorithm}

In the image reconstruction problem, the dirty image $\bm{I}^{dirty}$
can be expressed as
\begin{equation}\label{1}
\bm{I}^{dirty}=\bm{B \star I}^{true}+\bm{I}^{noise},
\end{equation}
where $\bm{B}$ is the PSF, $\bm{I}^{true}$ is the true image, $\bm{I}^{noise}$ is the noise convolved by the PSF and $\star$ is the convolution operator. One
goal of imaging algorithms is to remove the effects of the PSF. In practice, a model with a limited number of components is used
to approximate the true image,
\begin{equation}\label{2}
\bm{I}^{true}=\sum_{i=1}^{L}{\bm{I}_{i}^{component}}+\epsilon,
\end{equation}
where the model image $\bm{I}_{L}^{model} = \sum_{i=1}^{L}{
  \bm{I}_{i}^{component} }$, $\bm{I}_{i}^{component}$ is the $i$th
image component, $\epsilon$ is the error between the true image and the model image, and $L$ is the number of components used to approximate
the true image. In scale-insensitive algorithms, $\bm{I}_{i}^{component}$ is a delta function.
The true sky, including any extended emission, is thus modelled as a
collection of delta functions, which leads to residuals at levels much
higher than the thermal noise limit of modern telescopes, as shown by
various authors (including \cite{bha04}). The primary
limitation of the imaging performance of scale-insensitive algorithms is that their models cannot represent the correlation between pixels containing extended emission.  This is in conflict with the fundamental assumption underlying scale--insensitive algorithms that each pixel in the
image must be independent.  The Asp-Clean2004 algorithm tries to solve
the deconvolution problem by adaptively determining
the scales in various parts of the image by fitting the largest
possible scale locally using the following procedure:

\begin{enumerate}
\item Determine the initial values of the scale by smoothing the
  residual image with a few Gaussian beams, then identify the peak among
  the smoothed images and use the width of the
  Gaussian as the initial guess.
\item Refine the initial values with the LM minimization algorithm to
  determine the best model, $\bm{I}^{model}_{i} = \sum_{j=1}^{i} a_{j}e^{-\frac{1}{2}\frac{(x-x_{j})^{2}+(y-y_{j})^{2}}{\omega_{j}^{2}}}$, that fits the data locally.
\item Update the residual image as $\bm{I}_{i+1}^{residual}=\bm{I}^{dirty} - \bm{B}\star \bm{I}_{i}^{model}$.
\item Repeat steps 1--3 till one of the stopping criteria is satisfied. The stopping criteria can be the total number of iterations or a estimated noise threshold.
\end{enumerate}

As mentioned above, step 2 makes the component scales adaptive (i.e. they are not fixed and pre-determined), while the component scales are fixed and pre-determined in MS-Clean and its variants.
The minimization algorithm in step 2 minimizes the objective function $\chi^{2}$ , which is given by
\begin{equation}\label{3}
\chi^{2}=\|\bm{I}^{dirty}-\bm{B} \star \bm{I}_{i}^{model} \|_{2}^{2},
\end{equation}
where $\| \,\, \|_{2}$ is the Euclidean norm. In Eq.
\eqref{3}, there is a convolution operation between $\bm{B}$ and
$\bm{I}^{model}_{i}$. In a typical function minimization, the
objective function is evaluated many times for each fitted scale.  The FFT-based convolution used in the Asp-Clean2004
algorithm dominates the computational cost. For an $N-$pixel image, the
fast convolution includes two FFTs, one iFFT, and $N$ multiplication,
where the computational complexity of an FFT is about $N\log_{2} N$
\citep{gon10}. As such, the computational complexity of a convolution is
$N\left(3\log_{2} N+1\right)$ for an $N$-pixel image. For $M$
evaluations of the objective function, the computational complexity is
$MN\left(3\log_{2} N+1\right)$ when solving for a single component. While
the Asp-Clean2004 reconstruction gives good results, it is limited by the high computational cost.

\section{Asp-Clean2016: An efficient implementation of Asp-Clean2004}

As is well known, \st{the convolution of two Gaussian functions is another
Gaussian function. For example,} a two-dimensional Gaussian function
$g_{1}\left(a_{1},x_{1},y_{1},\omega_{1}\right)$ that is convolved with
another two-dimensional Gaussian function
$g_{2}\left(a_{2},x_{2},y_{2},\omega_{2}\right)$  results in a new
two-dimensional Gaussian function $g_{3}\left(a_{3},x_{3},y_{3},\omega_{3}\right)$, where
$a_{i},x_{i},y_{i}$ and $\omega_{i}$ are the amplitude, position
$(x_{i}, y_{i}),$ and width of Gaussian function $g_{i}$,
respectively. The Fourier convolution theorem states that if the
parameters of the convolved result (e.g. $g_{3}$) and a Gaussian
function (e.g. $g_{1}$) are known, the parameters of another Gaussian
function (e.g. $g_{2}$) can be analytically computed.

We therefore use the objective function
\begin{equation}\label{6}
\chi^{2}=\|\bm{I}_{i}^{residual}-\bm{I}_{ic}^{gauss}\left( a_{ic},x_{ic},y_{ic},\omega_{ic} \right)\|_{2}^{2},
\end{equation}
where $\bm{I}_{ic}^{gauss}$ is a Gaussian containing a model component whose
parameters (amplitude $a_{ic}$, location $x_{ic},y_{ic}$ and width $\omega_{ic}$) needs to be determined by
optimization (using the LM minimization).  After converging to the solution of
$\bm{I}_{ic}^{gauss}$, the underlying component that models the
emission is determined by analytically deconvolving the PSF
(approximated as a single Gaussian) as
\begin{equation}\label{7}
\omega_{i}=\sqrt{\omega_{ic}^{2}-\omega_{b}^{2} },
\end{equation}
where $\omega_{i}$, $\omega_{ic}$, $\omega_{b}$ are the widths of $\bm{I}_{i}^{component}$, $\bm{I}_{ic}^{gauss}$ and the main lobe of the PSF, respectively. The
amplitude $\alpha_{i}$ that corresponds to the $\bm{I}_{i}^{component}$
is calculated by the equation
\begin{equation}\label{8}
\alpha_{i}=\frac{\alpha_{ic} \omega_{ic}^{2}}{ 2 \pi \alpha_{b} \omega_{b}^{2} \omega_{i}^{2}},
\end{equation}
where $\alpha_{b}$ and $\alpha_{ic}$ are the amplitudes of the Gaussian beam approximated
from the PSF and $\bm{I}_{ic}^{gauss}$, respectively. As in other algorithms, a loop gain is used before the component $\bm{I}_{i}^{component}$ is added to the model image. In Eqs. \eqref{7} and \eqref{8}, the
$\alpha_{b}$ and $\omega_{b}$ of the PSF are fixed, and changes
of $\alpha_{i}$ and $\omega_{i}$ follow from any adjustments of
$\alpha_{ic}$ and $\omega_{ic}$.
Here we would like to point out similarities between our approximation
of the PSF by a single Gaussian (effectively ignoring all
levels of the PSF sidelobes) and use of an approximate PSF where only the
highest near sidelobe of the PSF is included in a similar step as in
Clark-Clean. The difference is that our algorithm uses this approximation to the
extreme in the estimation phase of the model components, while Clark-Clean uses it in the subtraction step in a minor cycle to
determine the next highest peak in the residual image.

Initial values are important to compute an optimal component. In many tests, we found that good initial values that are closer to
the optimal component can significantly reduce the iteration counts of
the minimization algorithm to find the optimal component. In other
words, good initial values can significantly reduce the computational time.
As in the Asp-Clean2004 algorithm, good initial values
can be obtained by smoothing the current residual image first and then
finding the parameters that correspond to the global peak among the
smoothed residual images. Smoothing is computed by a convolution, which has the effect of
increasing the computational cost. However, when deconvolving PSF-sized features, the peak of the last residual image and the scale of the main lobe of the PSF can be directly used as the initial values of the parameters. Then
the simple method can be triggered to reduce the computational cost. In the Asp-Clean2016 algorithm, the simple method is used only when
the area of the optimal component in the \textit{\textup{last}} iteration is smaller than $5\%$ of the area of the initial guess in the last\textup{\textup{\textit{}}} iteration. After this, the smoothing method is used again until the condition of using the simple method is met. In practice,
there are a large number of small scales in image decomposition. The modified scheme not only retains the reconstruction
quality, but also reduces the computational cost.

\section{Numerical experiment and comparison}

\begin{figure}[t]
\vspace*{0.2mm}
\begin{center}
\includegraphics[width=8cm]{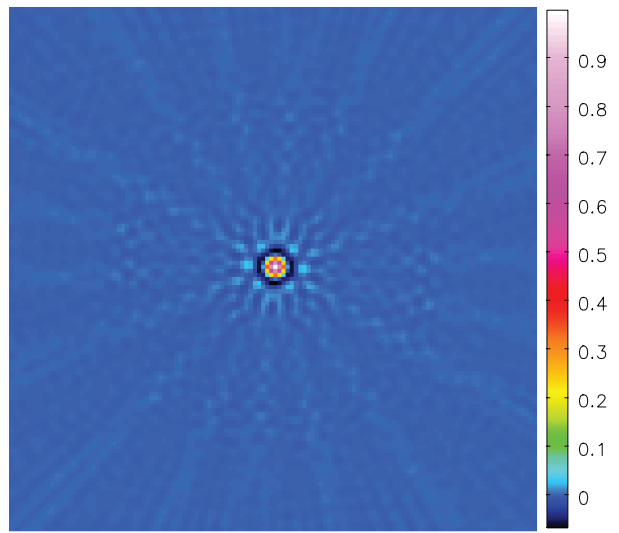}
\end{center}
\caption{Robust-weighted PSF in the range $-0.068$ to $1.0$ with the logarithmic scaling (CASA scaling power cycles = $-1.4$).}
\label{Dirty-B}
\end{figure}

We tested the Asp-Clean2016 algorithm and compared it with the
Asp-Clean2004 algorithm using simulated L-band EVLA data in the
B-configuration\footnote{http://www.aoc.nrao.edu/evla/}, with a bandwidth of $1$ GHz and $32$ channels. The observation time was six hours. All simulations
were made using the CASA software \footnote{http://casa.nrao.edu/}. The robust-weighted PSF is displayed in Fig.~\ref{Dirty-B}. The maximum negative sidelobe of the PSF was
$-0.068$ (the peak value of the PSF is normalized to $1$).  The full width at half maximum (FWHM) of the main lobe of
the PSF was about $2^{''}$, while the resolution of the images was $1^{''}$.

\begin{figure*}[t]
\vspace*{0.2mm}
\begin{center}
\includegraphics[width=16cm]{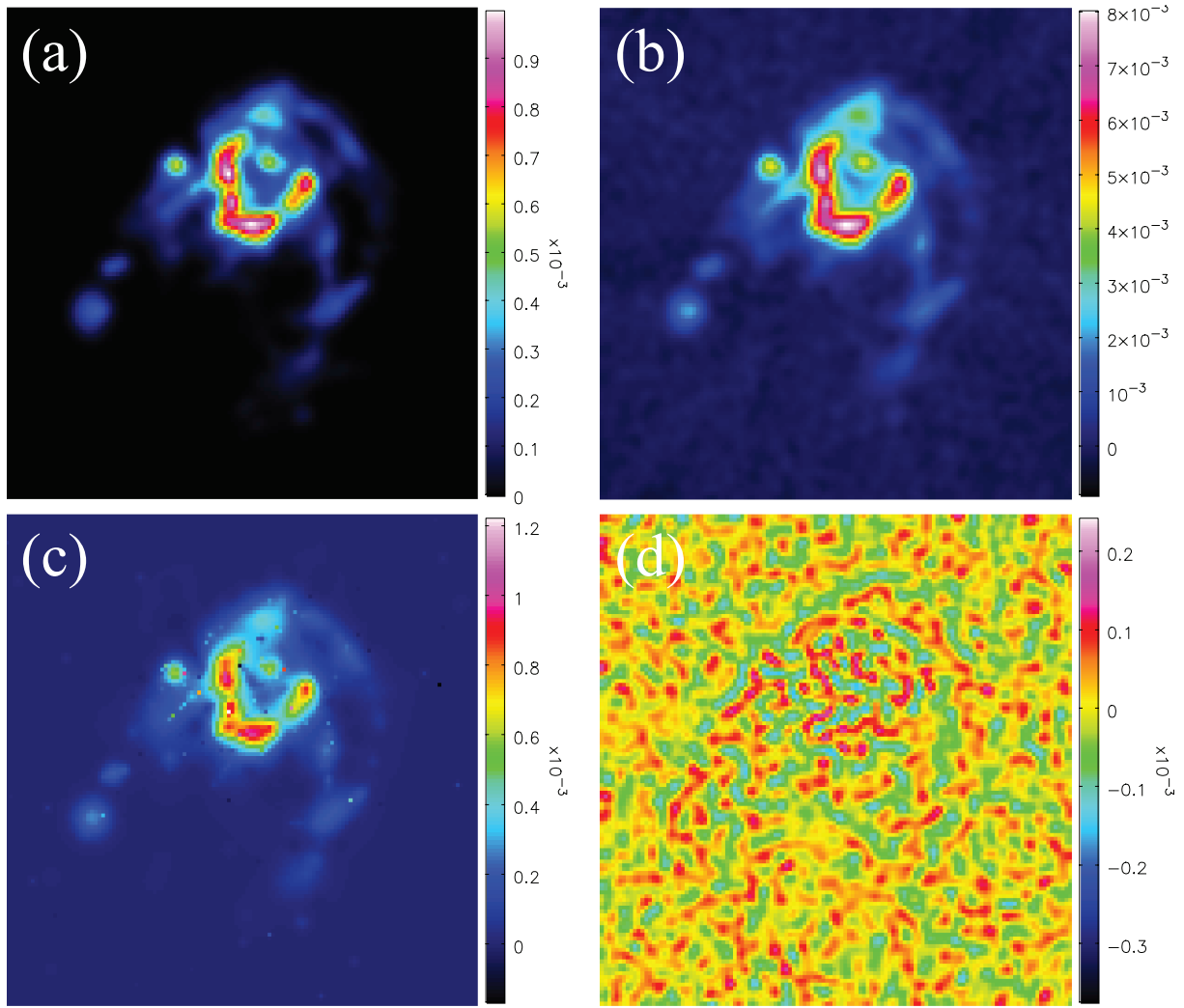}
\end{center}
\caption{Results of the M31 image from the Asp-Clean2016 algorithm: (a) the original image, (b) the dirty image with a robust weighting function, (c) the reconstructed model image, and (d) the corresponding residual image. }
\label{M31-IMG}
\end{figure*}

We tested the Asp-Clean2016
algorithm on the M31 image with robust weighting applied. The reconstructed results are displayed in
Fig.~\ref{M31-IMG}. The original image is shown in
Fig.~\ref{M31-IMG}(a), while the dirty image is displayed in
Fig.~\ref{M31-IMG}(b). Artificial Gaussian noise was added to the simulated visibilities, which resulted in a noise RMS of $5 \times 10^{-5}$ Jy in the dirty image. The loop gain used in the reconstruction was
$0.35$. Comparing the original image in Fig.~\ref{M31-IMG}(a), the
reconstructed image, which has about $900$ components (as shown in Fig.~\ref{M31-IMG}(c)),
can reconstruct various scales well.


\begin{figure*}[t]
\vspace*{0.2mm}
\begin{center}
\includegraphics[width=15cm]{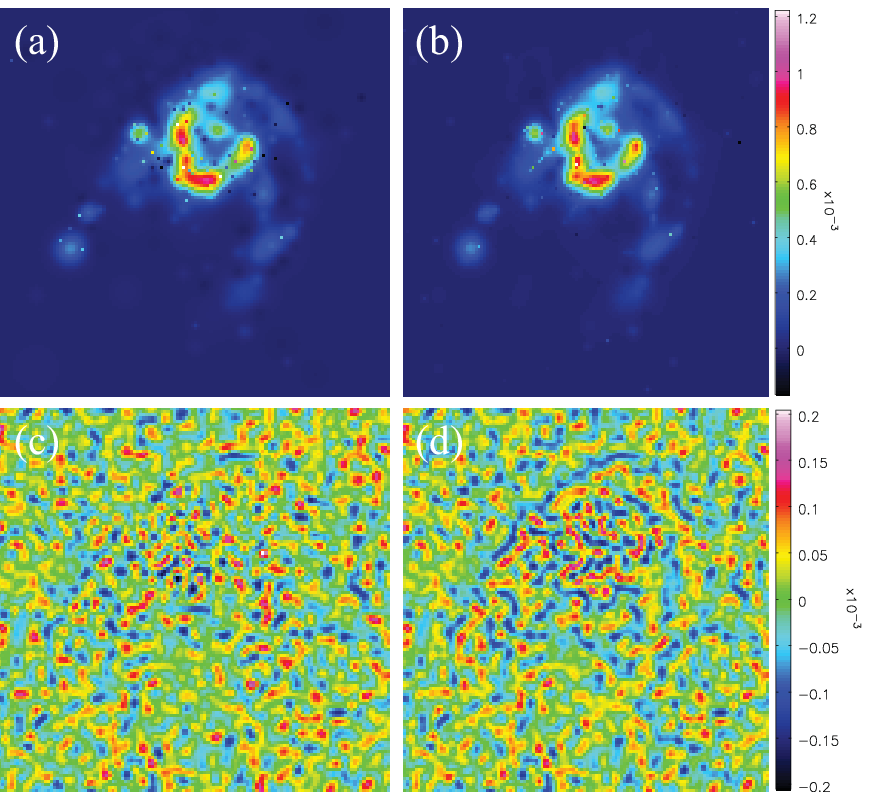}
\end{center}
\caption{Models and residuals of the M31 image from the Asp-Clean2004 and
  Asp-Clean2016 algorithms: (a) the Asp-clean2004 model image, (b) the
  Asp-Clean2016 model image, (c) the Asp-Clean2004 residual image, and
  (d) the Asp-Clean2016 residual image. The model images and the
  residual images are displayed in the same respective data ranges.}
\label{M31-COMP}
\end{figure*}

\begin{figure}[htbp]
\vspace*{0.2mm}
\begin{center}
\includegraphics[width=8cm]{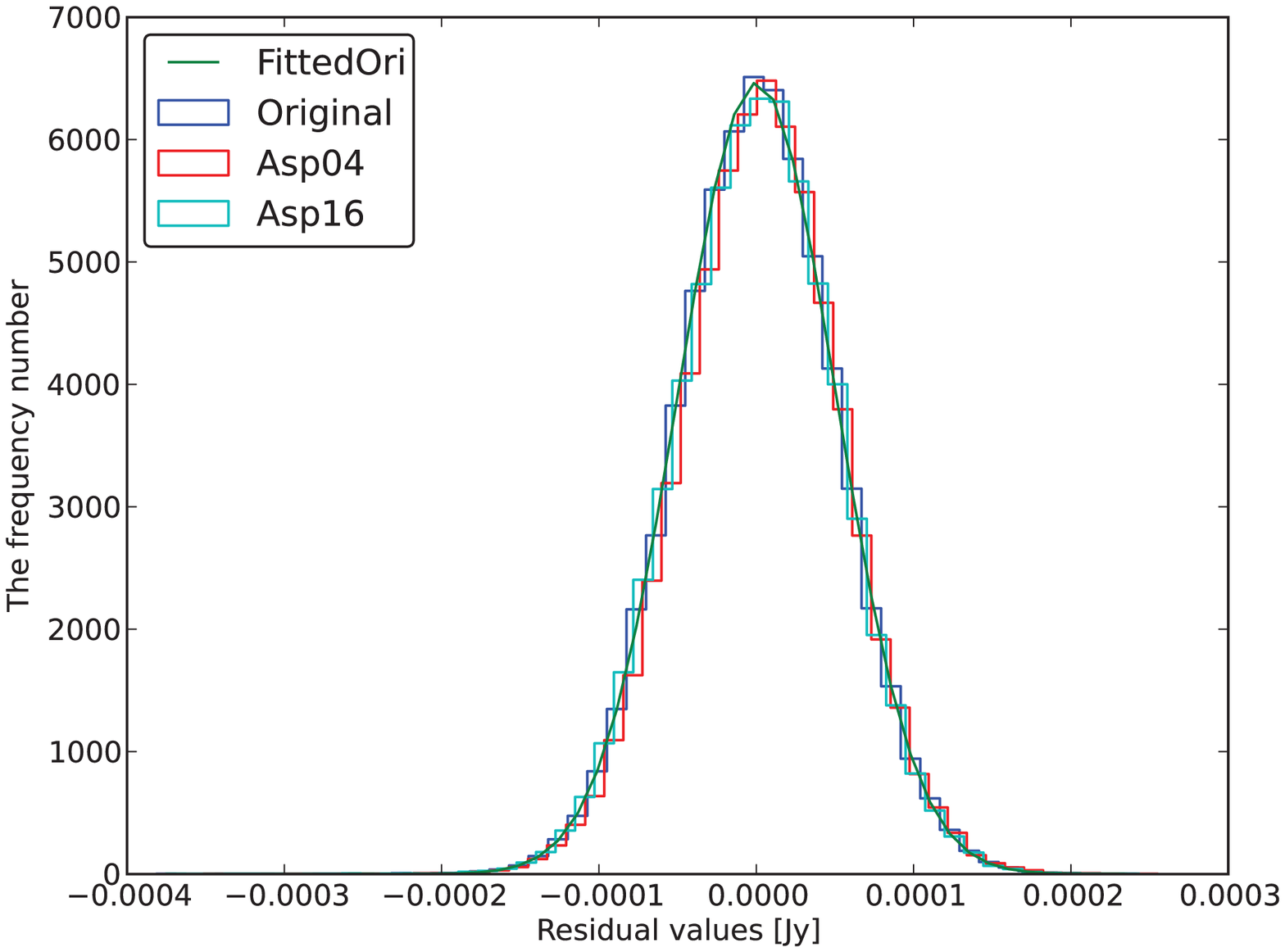}
\end{center}
\caption{Histogram of the residual images. ``FittedOri'' is a Gaussian fitted to the ``Original'' image noise. ``Original'' is the added image noise, which is the inverse Fourier transformation of the visibility noise. ``Asp04'' is the Asp-Clean2004 residual image, and ``Asp16'' is the Asp-Clean2016 residual image.}
\label{M31-RESI-COMP}
\end{figure}

We found that the reconstructed results of Asp-Clean2016
and Asp-Clean2004 are comparable (e.g. see Fig.~\ref{M31-COMP}). In Fig.~\ref{M31-COMP} the fundamental reason for the difference between the model (a) and (b) is that image reconstruction in interferometric imaging is non-unique. In other words, infinite numbers of image models will fit the data equally well. Two different algorithms will rarely produce the same set of image model components. However, they both should fit the data well. In Tables~\ref{TAB-1} and ~\ref{TAB-2} we show that the amplitude range and the total
reconstructed model flux from the Asp-Clean2016
algorithm is closer to the original image. However, the Asp-Clean2004
residual image is more noise-like and includes fewer correlated structures. From the histogram (see Fig.~\ref{M31-RESI-COMP}) of the residual images, we can see that the residual noise profiles of Asp-Clean2004 and Asp-Clean2016 are very close to each other and also very close to the input Gaussian noise. This shows that the new algorithm improved the computing efficiency without significant degradation in reconstruction quality. In all, the results of the Asp-Clean2016 results are
comparable to those of the Asp-Clean2004 algorithm, but the Asp-Clean2016 algorithm had a faster convergence rate. Table~\ref{TAB-3} shows that the Asp-Clean2016 algorithm converged about $20$ times faster in the test.


  \begin{table}[t]
   \caption{Model comparison between Asp-Clean2004 and Asp-Clean2016}
     \centering
\begin{tabular}
{c c c c}
\hline\hline 
 Models & Min           & Max           & Total Flux \\
 & ($10^{-4}$Jy) & ($10^{-4}$Jy) &  ($10^{-4}$Jy)\\
\hline 
\hline
Original  & $0.0$ & $10$ &  $6676$\\
\hline
Asp04-model & $-6.48$ & $13.87$ & $6167$\\
\hline
Asp16-model & $-1.67$ & $12.24$ & $6681$ \\
\hline
 \end{tabular}

  \label{TAB-1}
  \end{table}


  \begin{table}[t]
   \caption{Residual comparison between Asp-Clean2004 and Asp-Clean2016}
    \label{table:1}
     \centering
\begin{tabular}
{c c c c}
\hline\hline 
 Residuals & Min           & Max           & RMS\\
 & ($10^{-4}$Jy) & ($10^{-4}$Jy) & ($10^{-4}$Jy)\\
\hline 
\hline
Asp-Clean2004 & $-2.05$ & $2.05$ & $0.49$ \\
\hline
Asp-Clean2016 & $-1.56$ & $1.43$ & $0.51$\\
\hline
 \end{tabular}

  \label{TAB-2}
  \end{table}

To compare the performance of the Asp-Clean2004 algorithm and the Asp-Clean2016 algorithm with different weightings, we simulated the M31 image with uniform, natural, and robust weightings, respectively. The image sizes were the same: $256 \times 256$ pixels. The loop gain was $0.35$ and other parameters were the same. The numerical results are included in Table~\ref{TAB-3}. Different weightings in the spatial frequency domain result in different PSFs in the spatial domain, which in turn affects structures in the dirty image. However, the Asp-Clean2016 requires approximating the PSF as a Gaussian function. Obviously, the approximation is more effective when the PSF includes weaker sidelobes. Thus, for the Asp-Clean2016 algorithm, the convergence speed with a uniform weighting was faster than with a natural weighting under the same parameters. This is also partly because the performance of the minimization algorithm differs between dirty images. Table~\ref{TAB-3}
shows that the total runtime of Asp-Clean2004 is $20$ or more times longer than that of Asp-Clean2016. The mean one-iteration runtime ratios, which are defined as the ratio of total runtime and total numbers of iterations, are higher than $40$ in this test. In other words, the mean one-iteration runtime of Asp-Clean2004 is $40$ times longer than that of Asp-Clean2016. When reconstructing $512 \times 512$ M31 image, the Python code for Asp-Clean2016 took several minutes using a typical computer (Intel(R) Core(TM) i7-3770 CPU @3.4GHz,4.00 GB RAM). After it is implemented with C++, the runtime is expected to be much shorter.


  \begin{table}[t]
   \caption{Numerical comparison between the Asp-Clean2004 and Asp-Clean2016 algorithms}
    \label{table:1}
     \centering
\begin{tabular}
{c c c c c c}
\hline\hline 
 & Uniform & Natural  & Robust \\
\hline 
\hline
Total Runtime Ratio$\ast$  & 22.11 & 34.82  &  18.39 \\
\hline
One-iteration Runtime Ratio$\ast$ & 61.11 & 101.21  & 41.66 \\
\hline
Iteration Number Ratio$\ast$ & 0.36 & 0.34  & 0.44 \\
\hline

 \end{tabular}
  \tablefoot{$\ast$ The ratios are computed by dividing the quantities of the Asp-Clean2004 algorithm by that of the Asp-Clean2016 algorithm.}
  \label{TAB-3}
   \end{table}



\section{Summary}

 In this paper, we approximated the PSF with a single Gaussian, which
  allowed us to analytically compute the optimal components after a Gaussian was fitted to the current residual image. This greatly reduced the
  total runtime compared to the 2004 implementation of the Asp-Clean algorithm.
  We also showed through simulations that there was no significant
  degradation in the imaging performance when using such an
  approximate PSF. The approximation we made will only become
  less of a problem with data produced by future telescopes with $ \text{ten times}$ more antennas
  than current telescopes.  This will further reduce the PSF
  sidelobes, which justifies this
 analytical approximation of the PSF. In addition, we used a different scheme to determine the
initial values of the optimal Gaussian components, which were then passed
to the minimization algorithm.  In the 2004 implementation, initial values
were found by smoothing of the residual image with a few
Gaussians of different sizes.  The heuristic used in the 2016
implementation can switch between smoothing and the simple method to further
reduce the computational cost. Tests showed that Asp-Clean2016 converges faster than
Asp-Clean2004, especially for larger images.  The
current implementation of  Asp-Clean2016  is in Python and the CASA package. The Python source code is available \footnote{https://github.com/lizhangscience/}.
Implementing in the C++ in the CASA package is currently underway.

\begin{acknowledgements}
We would like to thank the people who develop Python and CASA, which provided an excellent development and simulation environment. Zhang L. thanks for support from the NRAO Graduate Student Internship program. The work was also supported by the National Basic Research Program of China (973 program: 2012CB821804 and 2015CB857100), the Nation Science Foundation of China (11103055) and the West Light Foundation of the Chinese Academy of Sciences (RCPY201105).
\end{acknowledgements}

\bibliographystyle{aa}

\begin{thebibliography}{}

\bibitem[Bhatnagar \& Cornwell(2004)]{bha04} Bhatnagar, S., \& Cornwell, T. J. 2004, A\&A, 426, 747
\bibitem[Clark(1980)]{cla80} Clark, B. G. 1980, A\&A, 89, 377
\bibitem[Cornwell(2008)]{cor08} Cornwell, T. J. 2008, EEE Journal of Selected Topics in Signal Processing, 2, 793    
\bibitem[Gonzalez \& Woods(2010)]{gon10} Gonzalez, R. C., \& Woods, R. E. 2010, Digital Image Processing, third edition   
\bibitem[H\"{o}gbom(1974)]{hog74} H\"{o}gbom, J.A. 1974, A\&AS, 15, 417
\bibitem[Marquardt(1963)]{mar63} Marquardt, D. W. 1963, Journal of the society for industrial and Applied Mathematics, 11, 431 
\bibitem[Rau \& Cornwell(2011)]{rau11} Rau, U., \& Cornwell, T. J. 2011, A\&A, 532, A71
\bibitem[Schwab \& Cotton(1983)]{sch83} Schwab, F. R., \& Cotton, W. D. 1983, AJ 88, 688
\bibitem[Thompson et al.(2001)]{tho01} Thompson, A. R.,Moran,J. M., \& Swenaon, G. W. 2001,Interferometry and synthesis in radio astronomy, 2nd Edition

\end{thebibliography}

\end{document}